# Nonlinear optical response and spontaneous polarization in layer-stacked gallenene using second harmonic generation

Muhammad Yunusa[1, †,*], Andrew K. Schulz[2,†,*], Tim Parker[3,†], Felix Schneider[3,†],
Kenan Elibol[4], Marius Predel[5], Jana Dzíbelová[5], Michel Rebmann[3], Taylan Gorkan[6], Peter A. van Aken[4],
Alfred J. Meixner[3], Engin Durgun[6], Jani Kotakoski[5], Dai Zhang[3, *], Metin Sitti[1,7*]

[1]Physical Intelligence Department, Max Planck Institute for Intelligent Systems, Stuttgart, Germany;
[2]Max Planck Institute for Intelligent Systems, Stuttgart, Germany;
[3]Institute for Physical and Theoretical Chemistry, University of Tübingen and LISA+, Tübingen, Germany;
[4]Max Planck Institute for Solid State Research, Stuttgart, Germany;
[5]University of Vienna, Faculty of Physics, Vienna, Austria;
[6]UNAM-National Nanotechnology Research Center and Institute of Materials Science and Nanotechnology, Bilkent University, Ankara 06800, Turkey
[7]School of Medicine and College of Engineering, Koç University, 34450, Istanbul, Turkey

*Corresponding author. Email: yunusa@is.mpg.de (M.Y.); aschulz@is.mpg.de (A.K.S.); dai.zhang@uni-tuebingen.de (D.Z.); sitti@is.mpg.de (M.S.)
†These authors contributed equally to this work.

**Gallenene is a promising low-dimensional material with a structure down to the thickness of a single atom, similar to graphene. However, van der Waals stacking of two-dimensional (2D) gallenene under confinement remain poorly understood. In this study, we present evidence of the formation of parallel-stacked hexagonal gallenene (a100) structures in liquid gallium. The present study demonstrates the AB stacking of 2D gallenene a100 crystals in liquid gallium sandwiched between two graphene layers, as observed through transmission electron microscopy. A nonlinear optical response of the confined hexagonal gallenene was investigated through second harmonic generation (SHG) microscopy. The SHG signal exhibits periodic peak intensity shifts upon angular rotation up to 90 degrees and intensity dampening at elevated temperatures. These findings offer insights on device applications of 2D gallenene**.

Liquid metals (LMs) employ the fluidic transport of liquids while simultaneously enabling substantial thermal and electrical conductivity, thereby facilitating a diverse range of applications spanning microelectronics to soft actuators (*1*). The extraordinary properties of gallium-based LMs are a result of their unique structure (*1*, *2*). LMs can be formed because of the supercooling effect in the pure form of gallium (Ga) when heated above its melting point (29.8 °C) or in an alloy composition. The typical Ga alloys are eutectic gallium-indium (75% Ga and 25% In by weight, 15.7 °C melting point) and gallium-indium-tin (68.5% Ga, 21.5% In, and 10% Sn by weight, -19 °C melting point) (*3*). Furthermore, the characteristics of gallium-based LMs are not limited to those of conventional liquids, as they also display exceptional properties such as anomalous density (*4*), phase separation (*5*), structural order (*6*), and chemical stability (*7*). Additionally, all LMs display exceptional thermal and electrical conductivity (*8*). In contrast to the majority of liquids, the high electrical conductivity of LMs enables their utilization as conductive inks (*9*, *10*).

The fundamental structure of LMs remains poorly understood despite their applications in engineering, chemistry, and physics. This is largely due to the opacity of LMs in their bulk state, which presents significant challenges to researchers. Low-dimensional structure studies of gallium metal under scanning tunneling microscopy have revealed, at the atomic level of resolution, the tendency to obtain stable single-indexed surface orientations of Ga(001), Ga(010), and Ga(100) up to the melting point of the bulk crystal. In these orientations, the (001) and (100) planes appeared as atomically flat terraces (*11*). The existence of stripe structures was demonstrated in gallium thin films grown on Si(111) via molecular beam epitaxy (*12*, *13*). Recently, the exfoliation of both gallenene (100) and (010) crystal planes from the liquid gallium phase has been reported on different substrates via solid-melt exfoliation technique (*14*). Through interaction of liquid gallium in between epitaxial graphene and SiC, single layer gallenene has been fabricated (*15*, *16*). Furthermore, epitaxial bilayer gallenene on GaN (0001) can be produced using metal organic chemical vapor deposition (*17*). A theoretical study has demonstrated that Ga can form stable monoatomic linear and zigzag chain structures that are metallic (*18*). Air-stable polar 2D metals (Ga and In) with out-of-plane polarization have been realized in hermetic seals through confinement by graphene layers in hetero-epitaxy technique (*19*, *20*). Additionally, ab initio molecular dynamics simulations have been carried out to investigate thermally stable gallenene layers (*21*, *22*). The concept of free-standing stacked layers of two-dimensional gallium crystals has been put forth (Fig. 1A), wherein self-assembled layer structures are capable of exhibiting positional and orientational order with typical electro-optical switching characteristics analogous to those observed in liquid crystal displays (LCDs) (*23*). To bridge the existing gap between the structural and functional understanding of supercooled liquid gallium (SLG) and its alloys, techniques such as transmission electron microscopy (TEM) and second harmonic generation (SHG), are essential for probing the internal structure of these liquids with precision.

SHG describes the phenomenon whereby a sinusoidal light wave of fundamental wavelength $\lambda$ is converted to a second harmonic wave with a wavelength of $\lambda/2$ upon interacting with a nonlinear medium (*24*). SHG imaging produces contrast based on the phase-matching conditions; if the phases are equal between the fundamental wave and the second harmonic wave, the SHG intensity increases; conversely, mismatched regions produce little to no SHG signal (*25*). Furthermore, SHG intensity is influenced by the excitation geometry, enabling the technique to be employed for example in the study of three-dimensional collagen structures within tissue (*26*, *27*) and the analysis of nanostructured metal compositions (*28*). In the case of metals, the SHG intensity is at its greatest when the excitation occurs along the crystal axis (*29*). This relationship has enabled the use of SHG for the study of the diverse compositions of metals, including gallium selenide (*30*), gold nanoparticles (*31*, *32*), transition-metal dichalcogenides (TMDC) (*33–35*), nematic liquid crystals (NLC) (*36*) and cadmium selenide nanowires (*37*).

In this report, we employ microstructural characterization techniques, including transmission electron microscopy and nuclear magnetic resonance spectroscopy, to elucidate the structure of naturally grown 2D gallenene flakes in confined LMs. The nanocrystal terminations under investigation are single-indexed gallium crystals with orientation a100, which have the potential to result in the formation of filaments (*38–40*). Subsequently, the nonlinear properties of the nanostructures were investigated through the utilization of nonlinear optical techniques based on SHG signal, with the objective of characterizing the structure order of these 2D stacked gallenene nanocrystals during angular, thermal and electrical perturbation. The nanocrystals are embedded in the confined SLG films, which are sandwiched between two indium-tin-oxide (ITO) conducting glass slides that are coated with a polyimide alignment layer. This layer is used in liquid crystals to align mesogenes (*41*, *42*). Furthermore, ferroelectric-like polarization switching in the confined SLG film was characterized via two-terminal device measurements.

We begin to study the nanocrystals of SLG (Fig. 1A) through optical microscopy, where we observed the formation of filament strands growing from the confined SLG (Fig. 1B) and optical micrograph of screw-like helical structures under reflection in linear cross-polarized light microscopy (PLM) (Fig. 1C). We visualized the broken helical filament structures with layers parallel to the substrates from the thin layer of SLG at room temperature in between polydimethylsiloxane substrate and acrylic-coated glass (fig. S1). Further microscopic analysis of exfoliated SLG on polyimide treated glass revealed the existence of large flat terraces topographically under PLM in transmission mode (Fig. 1D). Scanning electron microscopy (SEM) images of a fractured LM droplet slowly cooled under confinement show the clear arrangement of twisted filaments (figs. S2 and S3). To study the nanocrystal behavior at the atomic level, we encapsulated SLG by sandwiching it between two monolayer graphene sheets on a holey SiN TEM substrate (Fig. 1A). The TEM micrograph of the graphene-encapsulated gallenene provided evidence of multilayergallenene flakes with a100 orientation. For further verification, high-angle annular dark-field (HAADF) images of gallenene between the two graphene layers were taken (Fig. 1E). To confirm the presence of gallium and its metallic state, we performed electron energy loss spectroscopy (EELS). The obtained EELS spectra confirmed the presence of Ga-L edges (Fig. 1F). The SLG is not oxidized, as the O-K edge at 532 eV of the oxide was not observed (fig. S4).

To determine the atomic structure of the layered structures, we compared the atomic resolution HAADF image of gallenene (Fig. 1G) and the corresponding simulated HAADF image of 12 layer gallenene (Fig. 1H). From the high-resolution experimental results obtained, the a100 orientation of multilayer parallel stacked gallenene appears to be the stable termination on graphene. We compared the intensity profile recorded on the experimental and simulated HAADF images, which are in good agreement (Fig. 1I). The intensity profiles are recorded along the dashed green rectangle in Fig. 1, G and H. The fit of the results further confirms that the gallium structures are single-indexed with about 12 layer stacks of 2D gallenene, similar to previously reported studies for atomically flat terraces (*11*) or exfoliated monolayers (*14*). Moreover, the Fourier transform of the Gaussian filtered HAADF clearly shows the hexagonal ring structure of the gallenene a100 structure (fig. S5). The lattice spacing of the AB stacked gallenene a100 is measured to be $0.22 \pm 0.01$ nm, which agrees with the reported literature values of bilayer gallenene structure (*14*, *20*). Since gallium exhibits strong polymorphism with different crystal modifications, we investigated the possible dipole/quadrupole coupling of gallenene structures in SLG using nuclear magnetic resonance (NMR) spectroscopy (*43*). We observed a peak splitting at room temperature in capillary confinement (Fig. 1J). The $^{71}$Ga resonance frequency in the SLG is 122 MHz (*44*). The observed peak splitting is consistent with previous studies of liquid gallium in a confined porous matrix (*43*, *45*, *46*). In another view, the known polymorphism of gallium leads to different crystal modifications, where the initial condition of the liquid state plays a crucial role. The different precursors of the liquid state could contain quasi-clusters leading to the different components of the NMR signals (*45*, *46*).

Monolayer and stacked-layer gallenene structures were also investigated by density funcitonal theory (DFT) calculations (fig. S6). We obtained a relaxed atomistic model from monolayer to bilayer AB stacking. The bulk α-Ga structure crystallizes in the Cmca space group, with lattice constants from fully optimized structures determined as $a = 0.459$ nm, $b = 0.776$ nm, and $c = 0.459$ nm, which is in good agreement with experimental data (*14*). DFT analysis reveals that the a100-oriented gallenene monolayer has a planar hexagonal crystal structure with a Ga-Ga distance of 0.262 nm, consistent with previous reports (*14*, *47*). It displays metallic electronic characteristics (Fig. 2, A and B). The designed free-standing bilayer gallenene structure consists of AB-stacked planar a100 monolayers with bulk lattice parameters. Based on experimental data, only AB-type stacking was considered, with the interlayer distance set to the experimentally measured value of 0.22 nm. The electronic band structure of the bilayer a100 remains metallic (Fig. 2, C to D). Additionally, the trilayer a100 structure was analyzed with both ABA and ABC type stacking configurations, and the ab initio calculations indicate that ABA-type stacking is energetically more favorable (figs. S7, A to D). Moreover, the 12-layer structure was also modelled, considering both ABA...B and ABCA...C stacking configurations. The results indicate that the ABA...B stacking is energetically the most favorable (figs. S7, E to H).

Using a custom-built parabolic mirror confocal microscope (Fig. 3A) (*48*), we analyzed the SLG to observe the SHG through the gallium as a non-linear medium (fig. S8A), which allowed us to identify the amount of constructive or destructive interference (fig. S8B). We accomplished geometrical orientation changes of the supercooled liquid gallium sandwich-devices (SLG-SD) through imaging a static sample (Fig. 3B) with an angularly rotating beam electric field vector (Fig. 3, A to C). We saw a significant change in SHG intensity about the vector rotation angle, θ (Fig. 3D), indicating a favorable excitation geometry is present in the SLG (Fig. 3D). For individual spots, we observed a significant shift between aligned and perpendicular vector rotation, producing an up to 10.4-fold larger SHG signal for the aligned beam, indicating a preferred orientation due to a distinct geometry at a single spot (Fig. 3E and fig. S8A). This behavior indicates that the dipole moment of the fibers is parallel to the incident field (*35*). This trend is consistent with other materials, such as TMDC and ferroelectrics, which exhibit polar shifting consistency. This preferred orientation of excitation (figs. S8, A and B) is a useful mechanical feature because different parts of the SLG droplet have different excitation preferences, similar to that of other materials ranging from ferroelectrics (*49*, *50*) to TMDC (*33*–*35*), allowing tunable mechanical behaviors using the crystal geometry.

Upon further rotation of the electric field vector, we found that the SHG response increases again up to a second maximum of the SHG intensity at angles of 160-170 degrees (Fig. 3F), indicating an approximate 180-degree periodicity of the preferred orientation of the gallium fibers in the liquid metal (Fig. 1J). We saw that this periodic behavior is repeatable when we expanded to scanning to other spots (Fig. 3G, and figs. S9, B and C). This periodicity has been found in other transversely isotropic structures (figs. S8, A and B), including TMDCs (*34*, *35*) and ferroelectrics (*49*) highlighting the preferred orientations of these liquid gallium microstructures. In previous LM studies, a channeled glass plate has allowed the alignment of the LM crystals (*23*). If similarly possible in the liquid gallium, it would provide a preferred polar orientation to allow a uniform excitation geometry of the SLG, enabling the possibility of reliable switching between minimum and maximum SHG optical response using the electric field vector of the beam.

We then used the same custom-built confocal optical setup (Fig. 3A) to image the SLG-SD attached to a thermocouple device (Fig. 4A). We found a clear dependency of the SHG intensity on the temperature of the SLG sample (Fig. 4B). In a lower temperature range (23–80 °C), the SHG intensity increased steadily with a maximum found around 80 °C (Fig. 4, B and C). This assay is robust to cyclic temperature stress, and in this low temperature range, the physical structure of the sample appears to be unaffected. When the SLG reached the median range of temperatures (80 – 120 °C), we saw a steep drop in the SHG intensity, which decreased to 50% of the spectra recorded at room temperature (Fig. 4, B and C). Finally, at 140 °C, the SHG signal disappears completely for ten minutes. This behavior is accompanied by a drastic change in the scanned image of the sample (Fig. 4, B and C). Newly found spots in the scanned image emitted mostly luminescence, and the SHG signals obtained were very low or zero (Fig. 4B and fig. S10). This behavior indicates a phase change in the SLG, which has been previously described as a shift from the MP to the M2 and M1 phases in 1,3-dioxane (DIO) and SLG (*23*, *36*). We then cyclically decreased the temperature, resulting in the reappearance of the previously found scanned image and SHG signal. Although the SHG signal intensity is much weaker than for the same temperatures measured previously, it increases steadily until about 80 °C is reached. This is true for the entire temperature range (up to 200 °C) that is available for the thermocouple device. This may indicate that the SLG is rearranging itself to prevent further thermal damage. Higher temperatures result in a phase change behavior in the SLG. This temperature increase before the phase change affects the SHG signal, with the SHG signal disappearing after the phase change. Increasing the temperature further beyond this phase transition temperature leads to a weak reappearance of the SHG signal, indicating that the second phase, which has appeared due to structural rearrangement in the SLG at higher temperatures, also exhibits weak nonlinear optical properties. Similarly, when observed under linear cross-polarized optical microscopy, the intensity of the SLG texture reappeared above 120 °C during heating due to thermal agitation and structural rearrangement (*23*). This may indicate that the filaments in SLG maintain their structural order even at high temperature.

Finally, we imaged the SLG-SD with electrical bias (Fig. 4D) to test electrical perturbations using the same custom confocal SHG setup (Fig. 3A). We found that the susceptibility to electrical perturbation differs depending on the sample studied. In a few experiments, the physical structure of the sample changes significantly, which can be observed visually by the SHG scan of the stationary sample as the applied voltage is adjusted (fig. S11). Although these significant changes were observed with certainty, they can be restored to the previous state when the electrical perturbation is applied in the opposite bias, but in most cases not to the exact correlating extent of the electrical perturbation. This indicates that electrical perturbation can lead to significant structural changes in the liquid gallium. However, the observed changes in SHG are much smaller than those of thermal or geometrical changes in the structure, similar to that of DIO (*36*). This smaller effect on the electrical intensity could be due to the fact that the crystals in liquid Ga shift in their preferred orientation due to polarity. Therefore, after an electric charge is applied, they all experience an electric field force, but only some physically shift, as the electrical stimulation of the sample could be more localized compared to the thermal signal (*51*). Compared to the background, where the weakest signals are observed in the scanned image, spots of higher intensity are more sensitive to electrical perturbation (*52*), indicating that orientation changes are likely to occur.

We investigated SHG switching during electrical perturbations, which leads to optical switching of the SHG intensity, as shown in (Fig. 4, E to I). When the applied voltage was switched between 0 V and -10 V in multiple cycles, the SHG signal, on average, follows the trend of two presumed orientations switching repeatedly (Fig. 4F). The characteristics of this differ between spots, as some spots clearly show the optical switching (Fig. 4, G to I), while some spots only partially follow the trend. Similar results are reproduced when the applied voltage is switched between 0 V and +10 V (Fig. 4G) and between -10 V and +10V (Fig. 4I). The ability to consistently manipulate the response with polarity shifts allows for rapid electrical switching, which is a common use of liquid metals in microelectronics (*53*). Although all spots show individual behavior, the reliable switching trend of the average SHG intensity across all points of interest persists in all experimental cases, which could be due to variations in organization, structure, and order. In other words, we can control the magnitude of the intensity by perturbing the structural properties with positive and negative voltages.

Non-zero spontaneous electrical polarization in gallenene nanoflakes was observed in SLG. To characterize the spontaneous polarization, a triangular waveform method was used by applying a voltage across the cell (*54*). We measured the reversible polarization switching in a 50 μm thick cell sandwiched in between two ITO glasses (Fig. 4D), where the cell was prepared by shearing the SLG between the substrates. There are two possible cases to induce spontaneous polarization in the layered 2D crystals. The first is the spontaneous filament arrangement, and the second is the shear-induced layer sliding. We applied 22.5 $V_{pp}$ across the SLG film between the ITO glass electrodes at room temperature (Fig. 5A), while measuring the current through the cell. The applied electric field was 0.45 V/μm at a frequency of 0.5 Hz. The measured spontaneous polarization increases with the applied field strength when compared at different temperatures, where we showed that the change in spontaneous polarization occurs at 25 °C and 65 °C (0.2 Hz) (Fig. 5B). Furthermore, the dependence of the spontaneous polarization on the applied voltage frequency was investigated at different temperatures. We noticed a difference in the polarization with increasing temperature, and therefore, the non-zero spontaneous polarization could be maintained at temperatures higher than 160 °C (Fig. 5C). This nonzero polarization behavior of SLG is advantageous for nonvolatile memory application. In addition, the SLG cell exhibited a high and stable dielectric response even at an elevated temperature above 200 °C up from 0.1 MHz up to 10 MHz (Fig. 5D).

Metallic 2D layer stacks of gallium exist in their natural form under different conditions such as liquid confinement and cleavage approach. Our results provide details on the structure and nonlinear optical properties of the metallic 2D polymorph of gallium. We show AB stacking in 2D layers of gallium with hexagonal structure. The graphene-encapsulated structure exhibited hexagonal stacking. We have extended the use of modern and novel techniques such as SHG to probe the nonlinear activity of gallenene in SLG. The technique confirmed the sensitivity of the organized gallenene structures to external perturbations. These results are likely to motivate the studies on the metallicity and topological phenomena of 2D gallenene.

**Acknowledgments:** The authors thank Anitha Shiva for assistance with SEM imaging of graphene encapsulation, Amirreza Aghakhani for the help with impedance spectroscopy, Günter Majer and Igor Moudrakovski for NMR experiments. We thank Muazzam Idris for fruitful discussion on NMR analysis. We acknowledge Marko Burghard (MPI for Solid State Research, Stuttgart, Germany) for help with fabricating the graphene-encapsulated SLG samples. **Funding**: The Max Planck Society supported this work. This work was also supported by the German Research Foundation projects ME 1600/21-1, ZH 279/13-1 and ZH 279/16-1. **Author contributions**: MY and AKS conceived the project. MY and AKS fabricated devices and conducted electrical measurements. TP and FS performed SHG measurements. KE, MP, JD, and JK performed STEM analysis. TG and ED conducted DFT calculations. MY, AKS, TP, FS, MR, MP, JD, KE, and TG analyzed and interpreted the data. MY, AKS, JK, DZ, and MS supervised the research. MY, AKS, TP, FS, and KE wrote the manuscript with input from MR, MP, JD, TG, PAvA, AJM, ED, JK, DZ, and MS. All authors reviewed and commented on the manuscript. Data and materials availability: All data needed to evaluate the conclusions in the paper are present in the paper and/or the supplementary materials. Additional data related to this paper may be requested from the authors. **Competing interests:** Authors declare that they have no competing interests.

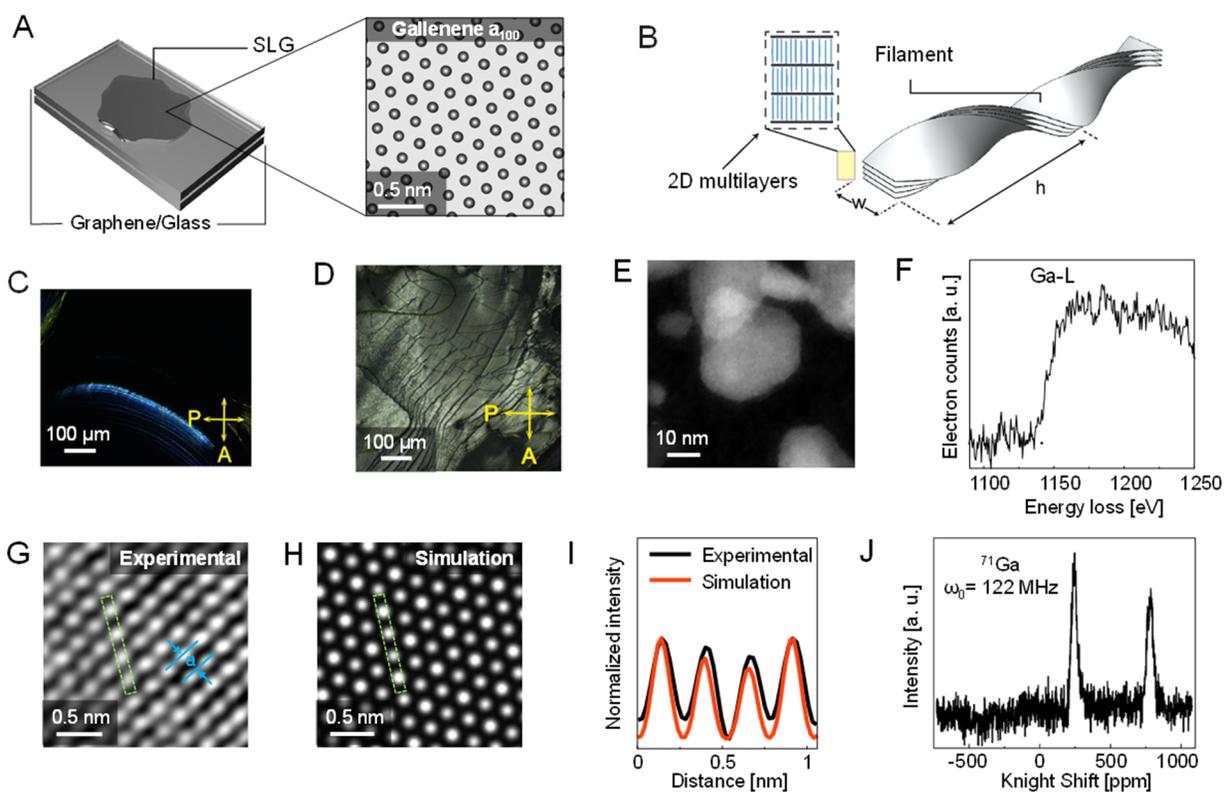

**Fig. 1. Structure of gallenene and complex anatomy of supercooled liquid gallium.** (**A**) Schematic of 2D gallenene structure in supercooled liquid gallium (SLG). (**B**) Filament model composed of twisted 2D multilayer of gallenene nanocrystals. The filament width is w and the helical half-pitch is h, respectively (**C**) Linearly cross-polarized optical micrograph of filament growing from the confined SLG thin film at room temperature. (**D**) Linearly cross-polarized optical micrograph of flat terraces of liquid exfoliated SLG. The yellow highlighted P and A, represent polarizer and analyzer (**E**) An example HAADF image of gallenene sandwiched between two graphene layers. (**F**) EELS spectrum obtained on the gallenene crystal shown in (E). (**G and H**) Atomic resolution HAADF image of multilayer gallenene and the corresponding simulated HAADF images of 12 layers, respectively. The lattice spacing is denoted as *a*. (**I**) Intensity profile recorded on the experimental and simulated HAADF images. The intensity profiles are recorded along the dashed green rectangle in (**G and H**). (**J**) $^{71}Ga$ NMR signal of SLG at room temperature. The peak splitting of the $^{71}Ga$ isotope is obtained in a micro-capillary filled with SLG. The resonance frequency $\omega_o$ of $^{71}Ga$ is 122 MHz.

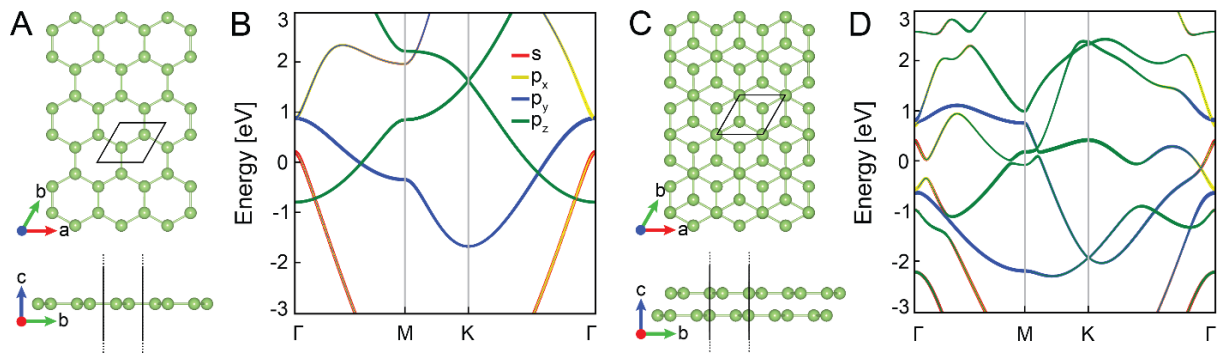

**Fig. 2. DFT calculations and band structure of gallenene.** (**A** and **C**) The top and side views of the optimized crystal structures of monolayer and bilayer a100. (**B**) The orbital projected band structure of monolayer a100, highlighting the Dirac cone at the K point above the Fermi level, which originates from the $p_z$ orbitals. The Dirac cones along the Γ-M or Γ-K directions, near the Fermi level, are formed by the $p_y$ band situated above the $p_z$ band. (**D**) The orbital projected band structure of bilayer a100, where the contribution near the Fermi level predominantly arises from $p_z$ orbitals.

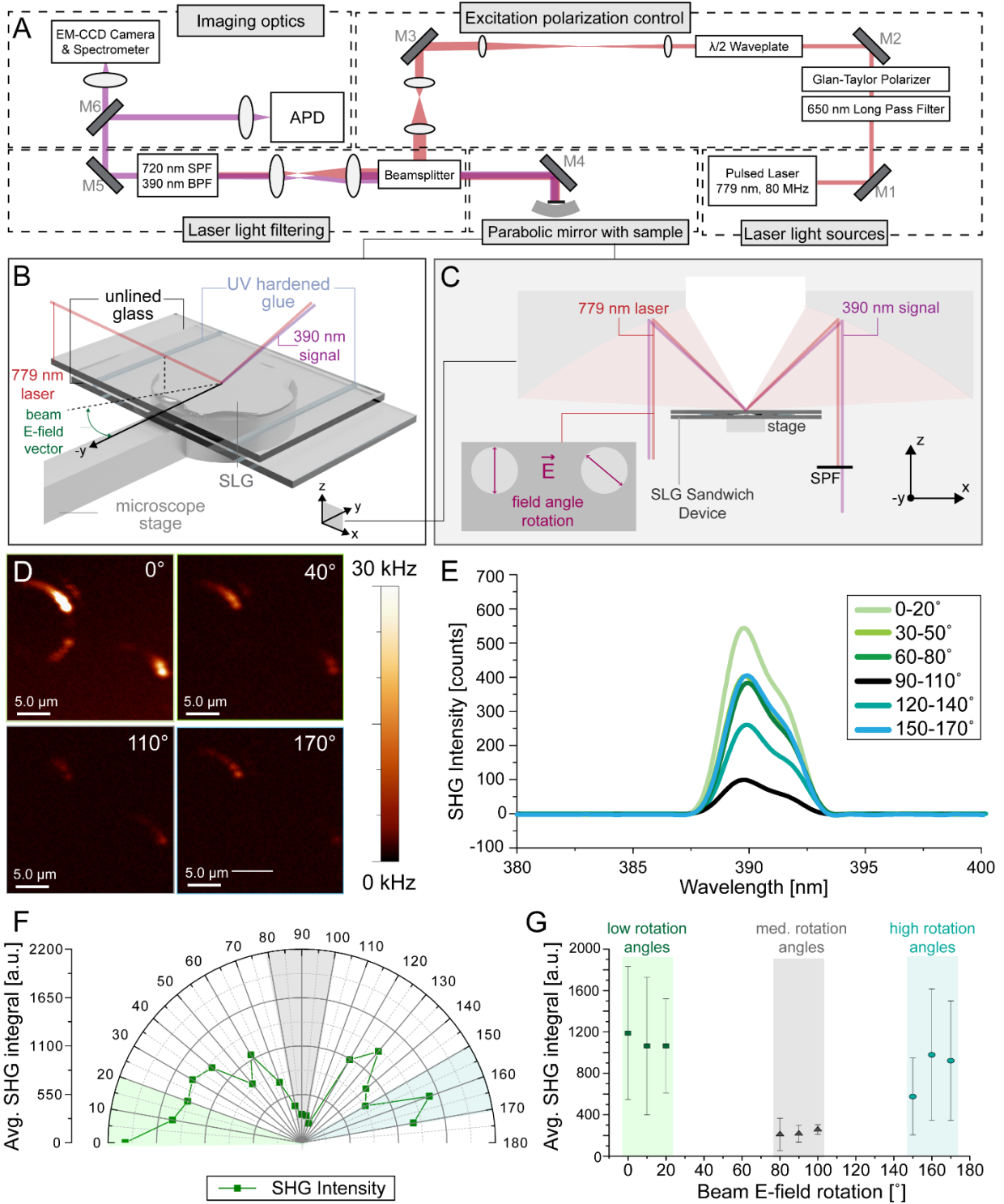

**Fig. 3. Custom-built SHG microscope and field angle rotation measurements.** (**A**) Custom-built confocal microscopy SHG setup with a static sample stage and a linearly polarized laser ($\lambda$= 779 nm, repetition rate = 80 MHz). It has five different sections, starting with laser light sources and ending with optics for imaging. (**B**) Schematic of the SLG in a sandwich device placed in the focus of the parabolic mirror from (**A**). (**C**) The configuration of sandwich device showing the field angle rotation. The angle of the incident electric field polarization is rotated via a $\lambda/2$ waveplate from (**A**). The SHG signal is separated from the fundamental laser light by a short-pass filter (SPF). (**D**)

Intensity images of the SLG at four different angles show the effect of different rotations on the SHG signal intensity. (**E**) Average over three points of the SHG of the SHG integral for different field angle rotations from 0 to 170 degrees. (**F**) Polar histogram showing the change in SHG intensity and the 170-degree periodicity of the maximum intensity of SHG signal in the SLG samples. (**G**) SHG intensities averaged over three spots that are excited with electric field of low, medium, and high rotation angles. The color codes for the lightly shaded bars are correlated with those in (**F**).

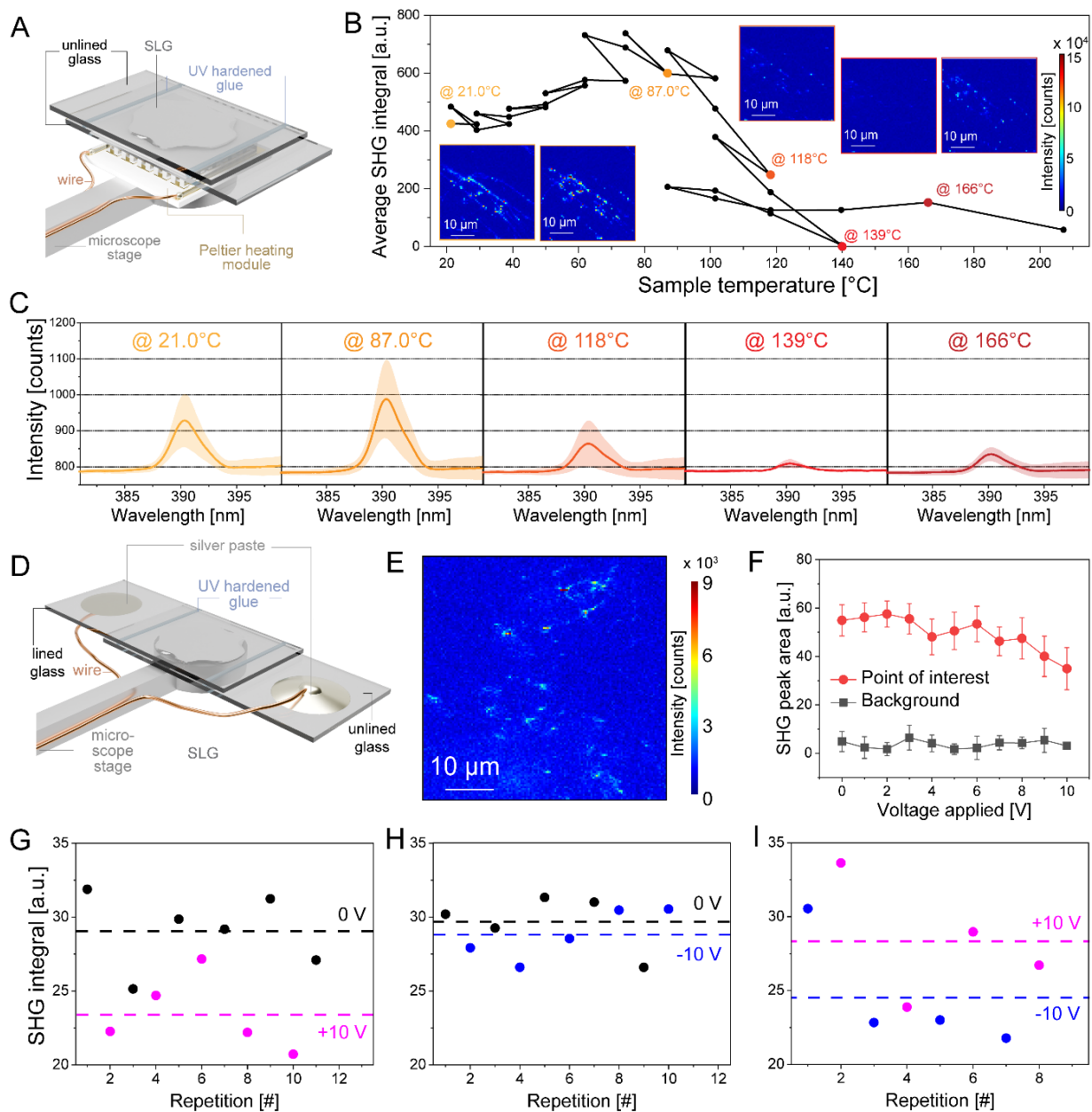

**Fig. 4. Thermal and electrical perturbations of SLG in SHG intensity. (A)** SHG intensity at different temperatures of an SLG sample on a Peltier hot plate, with the average SHG area shown in **(B)**. **(B)** Average SHG area over different sample temperatures with insets showing spectra of the SHG intensity starting from 20 °C up to 207 °C with insets showing five different spectra indicated by the yellow to red dots. (**C**) The average intensity at five highlighted points is shown in **(B),** with the intensity decreasing to nearly zero at 139 °C. **(D)** Schematic showing a new SLG sandwich device with wires for electrical charging. **(E)** Scan image of the SLG during electrical perturbation at +10 V. The applied voltage results in a visible deviation from the initial filament structure. **(F)** SHG intensity difference between background and the averaged points of interest during electrical perturbation. **(G and H)** SHG shifts between periodic repetitions with the average shown in dotted lines of **(G**) 0 V (black) and 10 V (pink), **(H):** 0 V (black) and -10 V (blue), and **(I)** +10 V (pink) and -10 V (blue).

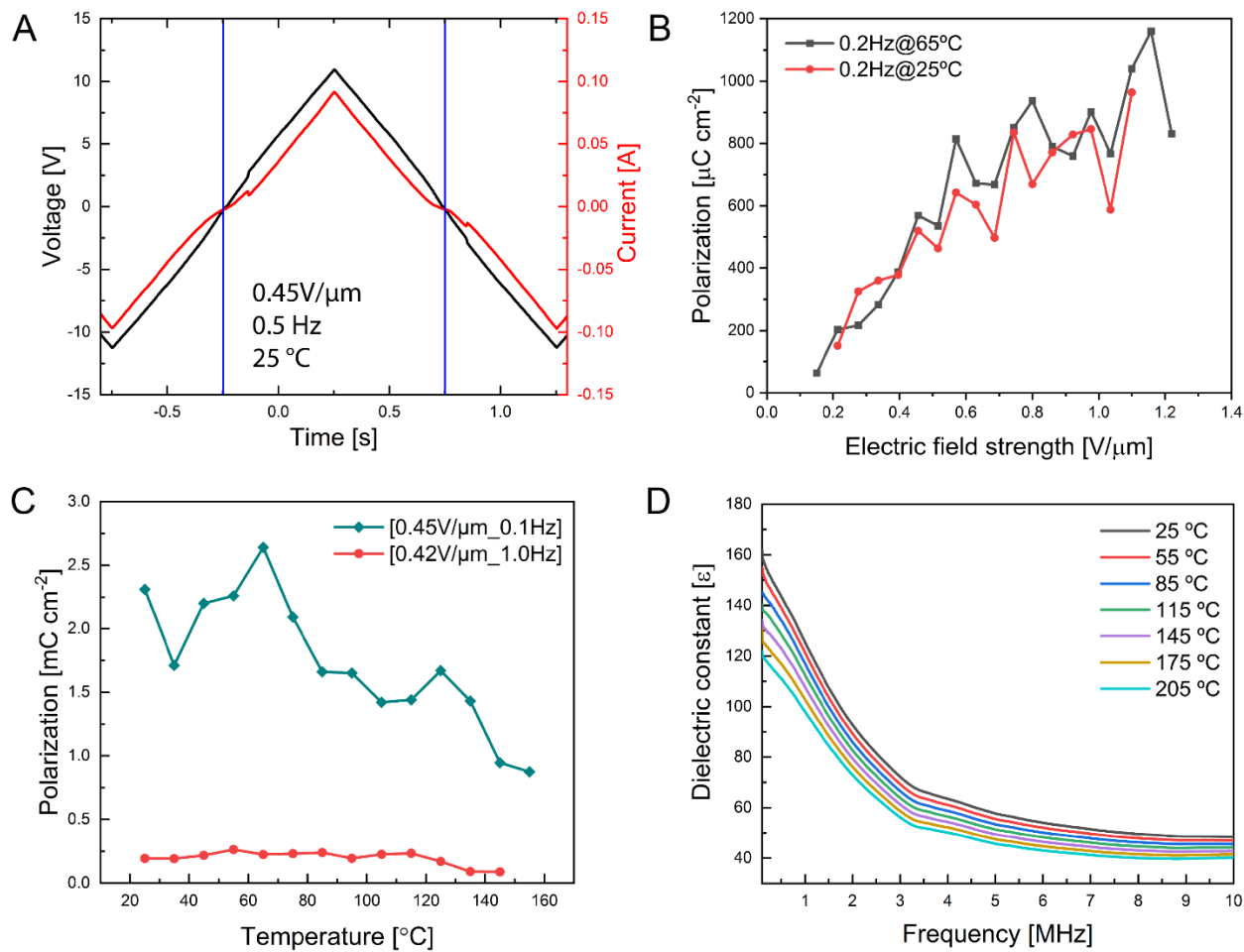

**Fig. 5. Spontaneous polarization, temperature and electric field response.** (**A**) Polarization current measurement at room temperature upon voltage with 22.5 V$_{pp}$ applied. (**B**) Electric field dependent polarization current measurements at 25 °C and 65 °C. (**C**) Comparison of polarization current with increasing temperature at 0.1 and 1 Hz. The temperature was increased at a rate of 5 °C/min and stabilized at the respective temperatures to measure the polarization current. Spontaneous polarization persisted even at high temperatures. (**D**) The dielectric response measured from room temperature up to 205 °C.